\documentclass[12pt]{iopart}

\usepackage{bm}
\usepackage{epsfig}
\usepackage{citesort}
\usepackage{graphicx}
\eqnobysec

\newcommand{\gwig}{\mbox{\;\raisebox{.3ex}
    {$>$}$\!\!\!\!\!$\raisebox{-.9ex}{$\sim$}}\;}
\newcommand{\lambdabar}{{\hbox{$\lambda$\kern-1.ex\raise+0.45ex\hbox{--}}}}

\begin{document}

\begin{flushright}
{\large \tt MPP-2007-19}
\end{flushright}

\title[Neutrino mass from future high redshift galaxy surveys]%
{Neutrino mass from future high redshift galaxy surveys: sensitivity and detection threshold}

\author{Steen~Hannestad}
\address{Department of Physics and Astronomy \\
University of Aarhus, DK-8000 Aarhus C, Denmark}

\author{Yvonne~Y.~Y.~Wong}
\address{Max-Planck-Institut f\"ur Physik (Werner-Heisenberg-Institut) \\
F\"ohringer Ring 6, D-80805 M\"unchen, Germany}

\ead{\mailto{sth@phys.au.dk},\mailto{ywong@mppmu.mpg.de}}

\begin{abstract}
We calculate the sensitivity of future cosmic microwave
background probes and large scale structure  measurements from galaxy redshift surveys
to the neutrino mass. We find that, for minimal models with few parameters, a
measurement of the matter power spectrum over a large range of redshifts
has more constraining power than a single measurement at  low
redshifts.  However, this improvement in sensitivity does not extend to larger models.
We also quantify how the non-Gaussian nature of the posterior distribution
 function with respect to the individual cosmological parameter
influences such quantities as the sensitivity and the detection
threshold. For realistic assumptions about future large scale
structure data, the minimum detectable neutrino mass at 95 \%
C.L.\ is about 0.05 eV in the context of a minimal 8-parameter
cosmological model. In a more general model framework,
however, the detection threshold can increase by as much as a factor
of three.
\end{abstract}
\maketitle


\section{Introduction}

Neutrinos differ from other types of energy density content
in our universe in two significant ways: They are essentially
non-interacting during the entire epoch of structure formation, and
transit from a relativistic to a non-relativistic particle species
also in the same epoch.
The first aspect means that the dynamics of neutrinos cannot be
described by simple fluid equations; effectively, the neutrino fluid
has infinite viscosity. The second aspect means that neutrinos
free-stream a finite distance before becoming non-relativistic; this
introduces a new scale, the neutrino free-streaming scale.

These unique features imply that cosmology can be used as a very efficient
tool to constrain the neutrino mass. For neutrinos of sub-eV masses,
the signature of massive neutrinos on the large scale structure (LSS)
power spectrum, i.e., the gradual damping of fluctuation power,
falls between wavenumbers $k\sim 0.01 \ h \ {\rm Mpc}^{-1}$ and
$0.5 \ h \ {\rm Mpc}^{-1}$, a range conveniently probed by current
cosmological probes.
Indeed, shape information from presently
available measurements of the LSS power spectrum, when
combined with observations of the cosmic microwave background (CMB) anisotropies, can already
constrain the upper limit on the sum of all neutrino mass
eigenstates $\sum m_\nu$ to lie in the range $0.2 \to 1 \ {\rm eV}$ in the context of
the concordance $\Lambda$CDM model. The precise value of the limit
depends strongly on the data sets used, as well as on the number of
free cosmological parameters
employed in the analysis (see
\cite{Goobar:2006xz,Seljak:2006bg,Feng:2006zj,Cirelli:2006kt,%
Hannestad:2006mi,Fogli:2006yq,Spergel:2006hy,Tegmark:2006az,Zunckel:2006mt,Kristiansen:2006ky}
 for a selection of recent papers on the
subject).

\begin{figure}
\hspace*{1.5cm}\includegraphics[width=100mm]{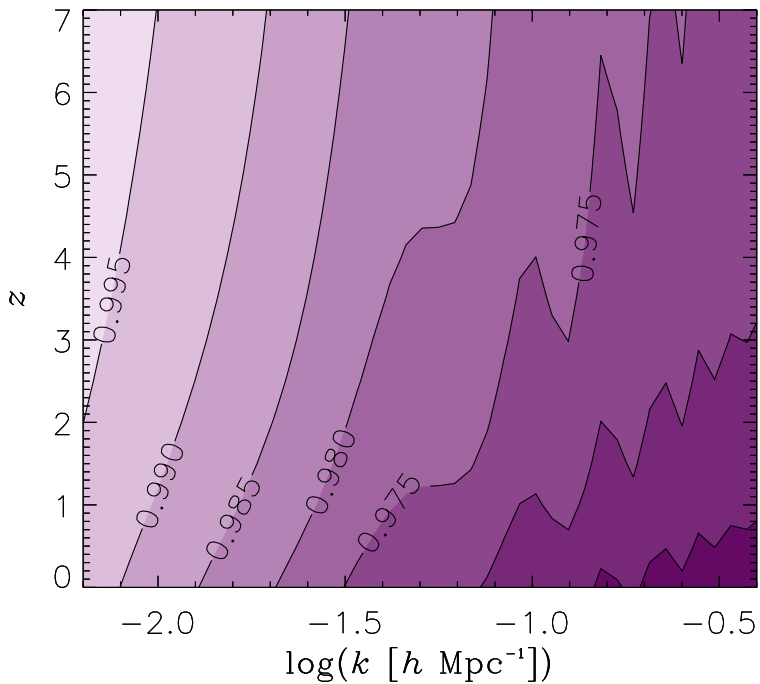} \caption{The
power spectrum suppression factor $P(k,z)/P_{m_\nu=0}(k,z)$ as a
function of wavenumber $k$ and redshift $z$ for a neutrino mass of
$\sum m_\nu = 0.08$ eV. The shadings are from 0.96 (darkest) to 1.00
(lightest) in steps of 0.005.} \label{fig:fig1}
\end{figure}

Furthermore, because the fraction of highly non-relativistic neutrinos changes
with time, and only highly non-relativistic neutrinos can cluster,
the impact of neutrinos on structure formation depends not only on scale, but also on time.
For neutrinos that are non-relativistic today, the LSS power
spectrum in terms of wavenumber $k$ and redshift $z$
is approximately given by \cite{Lesgourgues:2006nd}
\begin{eqnarray}
\hspace*{-2cm} P(k,z) \! = \! P_{\rm CDM}(k,0) \times
\cases{\left(\frac{g(z)}{(1+z)g(0)}\right)^2 & $k < k_{\rm fs}$ \cr
\left(\frac{g(z)}{(1+z)g(0)}\right)^{2-6f_\nu/5} \! (1-f_\nu)^3
[g(0)/a_{\rm nr}]^{-6 f_\nu/5} &  $k > k_{\rm fs}$}, \nonumber \\
\label{Eq:sup}
\end{eqnarray}
where $f_\nu = \Omega_\nu/\Omega_m$, $a_{\rm nr} = a(m_\nu=3
T_\nu)$ is the scale factor at which neutrinos become non-relativistic,
and $g(z)$ quantifies the decay rate of the gravitational potential.
Save for the case of $\Omega_{\rm total}=\Omega_m=1$,
$g(z)$ is a time-dependent function, so that the power spectrum suppression due to
neutrino free-streaming on scales below the free-streaming length
$\lambda_{\rm fs}=2 \pi/k_{\rm fs}$
is also time-dependent. Figure~\ref{fig:fig1} shows the
suppression in the case of $\sum m_\nu = 0.08 \ {\rm eV}$ relative
to $\sum m_\nu=0 \ {\rm eV}$
as a function of $k$ and $z$.
On large scales the two spectra are identical, while on small scales the suppression matches
equation (\ref{Eq:sup}).  The rapid oscillations in $P(k,z)/P_{m_\nu=0}(k,z)$
can be attributed to baryon acoustic oscillations (BAO) in the power spectra.

Clearly, cosmological probes that tap the power spectrum information
in terms of both the wavenumber $k$ and the redshift $z$
are {\it a priori} extremely
sensitive to the neutrino mass. One notable example is weak
gravitational lensing of distant galaxies, for which the impact of massive neutrinos has been
studied extensively (see, e.g.,
\cite{Cooray:1999rv,Abazajian:2002ck,Song:2003gg,Hannestad:2006as}).

Another example is high-redshift galaxy surveys.  The power of combining CMB
data with LSS measurements from low-redshift galaxy surveys to probe neutrino masses has, since the
pioneering work of \cite{Hu:1997mj}, been discussed
in many previous studies (see, e.g., \cite{Hannestad:2002cn,Lesgourgues:2004ps}).
In contrast, the advantage of using combinations of both low- and high-redshift surveys
has, apart from the study of \cite{Takada:2005si}, received relatively little attention so far.

The purpose of the present paper is to explore this last possibility
in detail.   A number of high-redshift galaxy surveys have been
proposed for the next decade and beyond.
The Wide Field Multiple Object Spectrograph
(WFMOS) survey proposed for the Subaru 8m telescope, for example, plans to
observe at redshifts $z\sim 1$ and $z \sim 3$ \cite{Glazebrook:2005ui}.
The Hobby-Eberly Telescope Dark Energy Experiment (HETDEX) will similarly look
out to $z \sim 4$ \cite{bib:hetdex}.  Further down in time, the
Cosmic Inflation Probe (CIP) mission  will observe at even higher
redshifts in space \cite{bib:cip}.

For simplicity we will adopt the survey set-ups used in \cite{Takada:2005si}.  However,
 our analysis differs from \cite{Takada:2005si} in several important ways:
(i) We use a simulation-based method for our parameter error forecast,
which has been shown to have many advantages over the popular Fisher matrix approach
\cite{Perotto:2006rj}.
(ii) We derive our constraints within more general cosmological frameworks that include uncertainties in,
e.g., dark energy equation of state, and other parameters degenerate with the neutrino mass.  This
has important implications for the sensitivity to $\sum m_\nu$ and its detection threshold.
(iii)  We opt to work with spherically-averaged, one-dimensional power spectra,
instead of two-dimensional spectra which in principle contain additional information from
geometrical and redshift-space distortions.
However, in contrast to the smoothed spectrum analysis of \cite{Takada:2005si},
we include also information from baryon acoustic oscillations.
Combination of geometrical/redshift effects and BAO
will likely lead to even more powerful parameter constraints
\cite{Seo:2003pu,Hu:2003ti}.
(iv) Several inconsistent assumptions in \cite{Takada:2005si}, particularly their
neglecting the effects of neutrino masses in the calculation of the CMB anisotropies, 
are rectified.

Lastly, let us note that, aside from the ability to probe the power spectrum evolution, going to high redshifts has the
advantage of reducing nonlinearities. At low redshifts ($z < 0.5$), uncertainties in the power spectrum
in the nonlinear regime require that we restrict the use of galaxy survey data to wavenumbers
not exceeding $k \sim 0.1 \ h \ {\rm Mpc}^{-1}$.
Indeed, for neutrino masses below $0.1 \ {\rm eV}$, the effects of nonlinearity
are already comparable to the damping of fluctuation power due to neutrino free-streaming
on scales around $0.05 \ h  \ {\rm Mpc}^{-1}$ at $z=0$.
At higher redshifts, however, nonlinear effects are less prominent,
and it is possible to probe safely the LSS power spectrum at much larger values of
$k$. This simple fact has important consequences for constraints on neutrino masses.

The paper is organised as follows.  In section~\ref{sec:models} we describe the
two fiducial models to be used in our investigation of the sensitivity of
future galaxy redshift surveys.
In section~\ref{sec:forecast} we describe the error forecast procedure: how mock 
galaxy power spectra are generated, the mock survey parameters we use, and the construction of 
the likelihood function.
We give in section~\ref{sec:results} the predicted sensitivities to various 
parameters in both cosmological models, and
we summarise our findings in section~\ref{sec:conclusions}.

\section{Fiducial models} \label{sec:models}

We consider two spatially flat cosmological models with eight and
ten free parameters respectively,
\begin{eqnarray}
{\bm \Theta}_{8} & = & \{\omega_b,\omega_c,\sum m_\nu,N_\nu,h,\tau,A_s,n_s\}, \nonumber \\
{\bm \Theta}_{10} & = & \{\omega_b,\omega_c,\sum m_\nu,N_\nu,h,\tau,A_s,n_s,\alpha_s,w\},
\end{eqnarray}
where $\omega_b \equiv \Omega_b h^2$ is the baryon density,
$\omega_c\equiv \Omega_c h^2$ the cold dark matter density, $\sum m_\nu$ the sum of neutrino masses,
$N_\nu$ the effective number of fermionic degrees of freedom during radiation
domination, $h$ the Hubble parameter, $\tau$ the optical depth to reionisation, 
$A_s$ the scalar perturbation amplitude, $n_s$ the
scalar spectral index, $\alpha_s$ the running of the scalar index
defined at a pivot scale of $k=0.05 \ {\rm Mpc}^{-1}$, and $w$
the dark energy equation of state parameter.  We consider two different 
models because nonstandard parameters such as $\alpha_s$ and $w$ are
known to be degenerate with $\sum m_\nu$ (e.g., \cite{Hannestad:2005gj}).
It is interesting to consider whether future galaxy redshift surveys are
actually able to break these degeneracies.

In our definition, $N_\nu$ enters the present-day energy density as
\begin{equation}
\label{eq:omeganu}
 \Omega_\nu h^2 =
 \frac{N_\nu m_\nu}{93 \, {\rm eV}}
 = \frac{\sum m_\nu}{93 \, {\rm eV}}.
\end{equation}
During the radiation-domination epoch the total energy density is
given by
\begin{equation}
\label{eq:rho}
 \rho = \frac{\pi^2}{30}\,T_\gamma^4
 \left[2 + 2 \times \frac{7}{8} N_\nu
 \left(\frac{T_\nu}{T_\gamma}\right)^4 \right],
\end{equation}
where $T_\gamma$ and $T_\nu$ are the photon and neutrino
temperatures respectively.  Our definition
of $N_\nu$ differs from that in \cite{Takada:2005si}: they use the 
definition (\ref{eq:omeganu}) for the present-day neutrino energy density,
but fix $N_\nu$ at $3.04$ in equation~(\ref{eq:rho}).

The dark energy density  $\Omega_\Lambda$ can be derived from the above parameters
using $\Omega_\Lambda=1-\Omega_\nu-\Omega_b-\Omega_c$.
The fiducial values of our model parameters are given in Table~\ref{table:fid}.

\begin{table}[t]
\caption{Fiducial values of the cosmological model parameters.\label{table:fid}}
\begin{indented}
\item[]\begin{tabular}{@{}lc}
\br
Model parameter & Fiducial value \\
\mr
$\omega_b$ &  0.0223 \\
$\omega_c$ & 0.105 \\
$\sum m_\nu$ & 0.0 \\
$N_\nu$  & 3.04 \\
$h$ & 0.7 \\
$\tau$ & 0.09 \\
$A_s$ & $2.3 \times 10^{-9}$ \\ 
$n_s$ & 0.95 \\
$\alpha_s$ & 0.0 \\
$w$ & $-1$ \\
$\Omega_\Lambda$ & 0.74 \\
\br
\end{tabular}
\end{indented}
\end{table}


\section{Mock data and parameter error forecast} \label{sec:forecast}

We use a simulation-based method to estimate cosmological parameter
errors from various combinations of future CMB and galaxy redshift
surveys.  Synthetic data are generated according to
the experimental specifications of the cosmological probe of interest,
and then analysed using a Markov Chain Monte Carlo (MCMC) package
such as CosmoMC \cite{Lewis:2002ah,cosmomc}.
The many advantages of MCMC-based forecasts over the popular
Fisher matrix approach are discussed in detail in \cite{Perotto:2006rj}.
Here, it suffices to reiterate that the MCMC method,
which probes the entire likelihood hypersurface,
generally provides more reliable results than does a Fisher matrix
analysis based on estimating the likelihood curvature around the best-fit
point.

Reference \cite{Perotto:2006rj} describes in detail an MCMC parameter
forecast using synthetic CMB data. We extend the analysis to include
also synthetic LSS data from galaxy redshift surveys as follows.

\subsection{Galaxy power spectrum}
\label{sec:mocksurveys}

Galaxy redshift surveys measure the correlation spectrum of
 galaxy number density fluctuations, $P_g(k)$.  In turn, this correlation
 spectrum is related to the underlying  matter power spectrum $P(k)$
via
\begin{equation}
P_g(k)=b^2 P(k),
\end{equation}
where the bias parameter $b$ varies according to the galaxy
type targeted by the survey at hand, but is generally expected to
be independent of $k$ in the linear regime.

Given some survey design and restricting the analysis to the linear
regime, one can expect to measure the galaxy power spectrum
$P_g(k)$ at a range of wavenumbers $k \in [k_{\rm min},k_{\rm max}]$
up to a statistical uncertainty of
\cite{Tegmark:1997rp}
\begin{equation}
\label{eq:error}
\Delta P_g(k) = \sqrt{\frac{1}{2 \pi  \ w(k) \ \Delta \ln k}} \left[P_g(k) + \frac{1}{\bar{n}_g} \right].
\end{equation}
Here, $w(k) = (k/2 \pi)^3 \ V_{\rm eff}$, and
\begin{equation}
V_{\rm eff} = \int d^3r \ \left[\frac{\bar{n}_g({\bm r}) P_g(k)}{1+\bar{n}_g({\bm r}) P_g(k)}
\right]^2
\end{equation}
is the effective volume of the survey, with $\bar{n}_g({\bm r})$ the
expectation value of the galaxy number density at coordinate ${\bm r}$.
When the condition $\bar{n}_g P_g(k) \gg 1$ holds,
$V_{\rm eff}$ is equivalent to the actual volume of the survey.

The quantity $\Delta \ln k$ is the bin size at $k$ in $\ln k$-space.
In general we use a very fine binning scheme to capture as much of the features in
$P_g(k)$ as possible, including wiggles from baryon acoustic oscillations
(BAO) at $k \gwig 0.01 \ {\rm Mpc}^{-1}$.
The effect of smearing due to finite window functions,
and hence a partial loss of the BAO features in a realistically
reconstructed power spectrum,
 will be discussed in
section \ref{sec:windowfunctions}.

\begin{table*}[t]
\caption{Mock galaxy survey specifications.  From left to right, $z_c$ is the
central redshift of each survey slice, $k_{\rm min}$ ($k_{\rm max}$)
the minimum (maximum) wavenumber probed by the slice, $V_{\rm eff}$ the slice's effective
volume, $\bar{n}_g$ the comoving number density of galaxies, and $b$ the
bias factor.  See \cite{Takada:2005si} for a discussion of the calculation of $b$.
\label{table:surveys}}
{\footnotesize
\hspace{12mm}
\begin{tabular}{@{}lcccccc}
\br
Survey & $z_c$ & $k_{\rm min}$ & $k_{\rm max}$
& $V_{\rm eff}$  & $\bar{n}_g$  & $b$  \\
& &  [$10^{-3} \ h \ {\rm Mpc}^{-1}$]  & [$h \ {\rm Mpc}^{-1}$]  & [$h^{-3} \ {\rm Gpc}^{3}$] &
[$10^{-3} h^3 \ {\rm Mpc}^{-3}$] & \\
\br
G1  & 0.75 & 5.32 & 0.14 & 1.65 & 0.5 & 1.22 \\
 ($0.5 < z < 2$) & 1.25 & 4.54 & 0.19 & 2.65 & 0.5 & 1.47 \\
        & 1.75 & 4.26 & 0.25 & 3.20 & 0.5 & 1.75 \\
\mr
G2  & 2.25 & 7.15 & 0.32 & 0.68 & 0.5   & 2.03 \\
 ($2 < z < 4$)  & 2.75 & 7.11 & 0.41 & 0.69 & 0.5  & 2.32 \\
    & 3.25 & 7.18 & 0.52 & 0.67 & 0.5  & 2.62 \\
    & 3.75 & 7.29 & 0.64 & 0.64 & 0.5  & 2.92 \\
\mr
SG  & 4 & 5.82 & 0.71 & 1.26 & 5 & 4 \\
($4 < z < 6$)   & 5 & 6.03 & 1.01 & 1.13 & 5  & 5 \\
    & 6 & 6.24 & 1.50 & 1.02 & 5  & 5.5 \\
\br
\end{tabular}
}
\end{table*}

\subsection{Mock survey parameters}

Following \cite{Takada:2005si}, we consider galaxy surveys probing three
different ranges of redshifts:
\begin{itemize}
\item G1: $0.5 < z < 2$, ground-based, $\Omega_{\rm sky} = 1500 \ {\rm deg}^2$,
\item G2 : $2 < z < 4$, ground-based, $\Omega_{\rm sky} = 300 \ {\rm deg}^2$, and
\item SG : $4< z< 6$, space-based, $\Omega_{\rm sky} = 300 \ {\rm deg}^2$,
\end{itemize}
where $\Omega_{\rm sky}$ denotes the survey's sky coverage.
Note that our G1 survey
has five times the sky coverage (and hence survey volume) of the survey of the same name
in \cite{Takada:2005si}.
We further subdivide each survey by redshift, although we have also checked that,
within our one-dimensional (i.e., spherically-averaged spectrum) approach, subdivision or not makes
essentially no difference to the final results.
Table \ref{table:surveys} shows the
central redshift $z_c$, effective volume $V_{\rm eff}$, and
other specifications for
each of these mock surveys and their subdivisions.
As a rough guideline, the specifications of G1 are similar to those of the
$z \sim 1$ WFMOS survey \cite{Glazebrook:2005ui}, while G2 is akin to
HETDEX \cite{bib:hetdex} or  the $z \sim 3$ WFMOS survey \cite{Glazebrook:2005ui}.
The space-based survey SG should be a reasonable approximation of CIP \cite{bib:cip}.
See \cite{Takada:2005si} for a more detailed discussion.

Note that the range of wavenumbers probed varies from survey to survey,
and, indeed, from redshift slice to redshift slice.  At the low end of the spectrum,
the slice/survey volume defines the minimum wavenumber $k_{\rm min}$ available for
observation,
\begin{equation}
k_{\rm min} = 2 \pi/V_{\rm eff}^{1/3},
\end{equation}
so that the largest observable perturbation wavelength  does not
exceed the slice's effective length scale.  At the other extreme,
the maximum wavenumber $k_{\rm max}$ is chosen according to the
criterion that the dimensionless power spectrum,
\begin{equation}
\Delta^2(k) \equiv \frac{k^3 P(k)}{2 \pi^2},
\end{equation}
does not exceed unity in the linear theory.  Rather than setting
common $k_{\rm min}$ and $k_{\rm max}$ for all surveys/slices of
interest, we choose to adhere to these ``natural'' limits, because
these limits reflect better the true strengths and drawbacks of 
each survey set-up.

 \subsection{Mock power spectrum data and the likelihood function}

The first step of an MCMC forecast consists of generating a vector of
$N$ observed data points
\begin{equation}
\bm{d} = \left( \begin{array}{ccc}
        P^{\rm obs}_g(k_1), &
        \cdots, &
        P^{\rm obs}_g(k_N) \end{array} \right)^T,
\end{equation}
and the corresponding error covariance matrix
\begin{equation}
\bm{N}= {\rm diag} \left( \begin{array}{ccc}
            [\Delta P_g(k_1)]^2, &
            \cdots, &
             [\Delta P_g(k_N)]^2 \end{array}\right),
\end{equation}
for a given fiducial model and survey/slice. For simplicity, we take
the data points $P^{\rm obs}_g(k_i)$ to be equal to the
fiducial power spectrum $P_g^{\rm fid}(k_i)$, with errors given by
equation (\ref{eq:error}).  A more rigorous analysis would require
that we draw a random $P_g^{\rm obs}(k_i)$ from a Gaussian centered
on $P_g^{\rm fid}(k_i)$ with variance $[\Delta P_g(k_i)]^2$ at every
$k_i$, but we choose not to pursue this avenue.

Since both the signal and noise are Gaussian-distributed,
we can construct a likelihood function in the form
\begin{equation}
\label{eq:likelihood}
L  \propto \exp \left[- \frac{1}{2} (b^2 {\bm v} - {\bm d})^T {\bm N}^{-1} (b^2 {\bm v} - {\bm d}) \right],
\end{equation}
where the vector ${\bm v}$,
\begin{equation}
\bm{v} = \left( \begin{array}{ccc}
        P^{\rm th}(k_1), &
        \cdots, &
        P^{\rm th}(k_N) \end{array} \right)^T,
\end{equation}
denotes theoretical predictions of the matter power spectrum $P(k)$.
Since the $N$ data points are assumed to be uncorrelated, the likelihood function
is equivalent to
\begin{equation}
\chi^2 \equiv - 2 \ln L = \sum_i^N \left[\frac{P^{\rm obs}_g(k_i)-b^2 P^{\rm th}(k_i)}
{\Delta P_g(k_i)}\right]^2,
\end{equation}
up to a constant offset.

Observe how the bias parameter $b$ enters the likelihood function
(\ref{eq:likelihood}).
In general we are not interested in the exact value of $b$; instead,
we construct an effective likelihood function $L_{\rm eff}$ by
marginalising over $b^2$,
\begin{equation}
L_{\rm eff} \propto \int  d b^2 \ \pi(b^2) \ L.
\end{equation}
If the prior $\pi(b^2)$ is flat, the marginalisation can be performed analytically
to give
\begin{eqnarray}
\chi^2_{\rm eff} &\equiv& -2 \ln L_{\rm eff} \nonumber \\
&=& {\bm d}^T \left({\bm N}^{-1} - \frac{{\bm N}^{-1} {\bm v} {\bm v}^T {\bm N}^{-1}}
    {{\bm v}^T {\bm N}^{-1} {\bm v}} \right) {\bm d} + \ln({\bm v}^T {\bm N}^{-1} {\bm v}),
\end{eqnarray}
again, up to a constant offset.

For multiple surveys/slices, the combined likelihood is simply a product
of the individual effective likelihood functions.  The CMB likelihood can
be also be incorporated in a similar fashion.

\subsection{Window functions}
\label{sec:windowfunctions}

Our treatment so far supposes that the power spectrum, including
the BAO wiggles, is well-sampled.  In reality, however, the finite
widths of the window functions used in the reconstruction of the
power spectrum from a galaxy survey will smear out the power in
Fourier space, so that little of the BAO features remain in the
reconstructed spectrum;  A separate analysis in terms of real
space two-point correlation must be performed in order to
extract the BAO peak.

It is interesting to compare the sensitivities of future galaxy
redshift surveys both with and without BAO extraction.
As said, the former case is already covered by our default treatment.
For the latter case, we mimic the effect of
a finite window function by smoothing the fiducial
power spectrum with a normalised top-hat function in log space,
\begin{equation}
\ln P^{\rm sm}_g(k) = \int d \ln k' \  W(k,k',R) \ \ln P_g(k'),
\end{equation}
where
\begin{equation}
W(k,k',R) = \left\{ \begin{array}{ll}
        1/R, & \qquad \ln k-R/2 \leq \ln k' \leq \ln k+R/2, \\
        0, & \qquad {\rm otherwise.} \end{array} \right.
\end{equation}
We use $R  = 0.33$, roughly similar to the ``$1 \sigma$''
width of the SDSS LRG window functions \cite{Percival:2006gs}.  Mock data points are then drawn
from the smoothed spectrum $P^{\rm sm}_g(k)$ with errors given by (\ref{eq:error}).

\subsection{Mock CMB data}

Since the LSS power spectrum is generally fairly featureless,
parameter degeneracies abound so that measurements from galaxy redshift
surveys alone do not place very stringent limits on cosmological parameters.
In order to break these degeneracies, it is common
to consider power spectrum measurements together with data from CMB
observations.  In the present work, we consider prospective CMB data
from the Planck satellite \cite{bib:planck}.  We assume measurements of the auto and
cross correlation power spectra of the CMB temperature and $E$-type polarisation
(i.e., TT, TE, EE), up to a multipole $\ell = 2250$.  Experimental characteristics of
the Planck satellite, assuming one year of observation,
can be found in Table 1.1 of the Planck Bluebook \cite{bib:bluebook}.

\begin{figure}
\hspace*{2.5cm}\includegraphics[width=130mm]{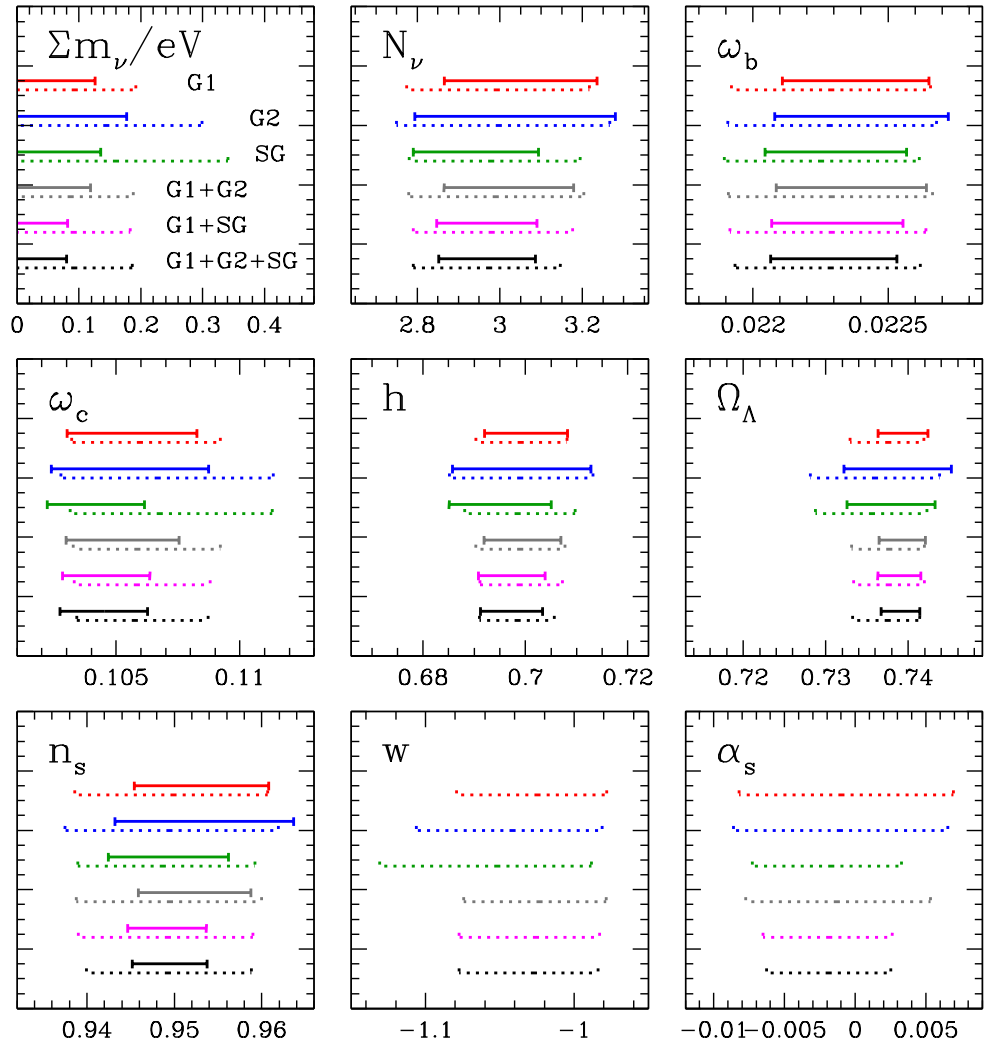} \caption{
Projected 1D marginalised 95 \% confidence regions for various
cosmological parameters within the 8- and 10-parameter models, using
various combinations of Planck and mock galaxy surveys.  In each
subplot, the data sets under consideration are, from top to bottom,
Planck+G1, Planck+G2, Planck+SG, Planck+G1+G2, Planck+G1+SG, 
and Planck+G1+G2+SG.
Solid lines denote constraints in the context of the 8-parameter
model, while dotted lines refer to the 10-parameter model.
 \label{fig:sigma} }
\end{figure}

\section{Results} \label{sec:results}

Figure~\ref{fig:sigma} shows the projected 1D marginalised 95 \% confidence regions
for various parameters within our 8- and 10-parameter models, assuming data from
Planck plus different combinations of mock galaxy surveys.

Consider first the 8-parameter model.
We note that there is a marked improvement in the projected bounds on
$\sum m_\nu$ when galaxy survey data from G1 ($0.5 < z < 2$) and SG ($4 < z < 6$)
are combined.  This is particularly interesting because the G1 survey has more than
twice the effective volume of the SG survey.   However, since SG is able to access a higher
$k_{\rm max}$ and, being a space-based probe, also suffers from less shot noise,
the Planck+G1 and Planck+SG bounds on $\sum m_\nu$ are almost identical: 
$\sum m_\nu< 0.13 \ {\rm eV}$ (95 \% C.L.).
Furthermore, the fact that the underlying LSS spectrum is both scale- and time-dependent
(as demonstrated in Figure~\ref{fig:fig1}) means that the G1 and SG mock data sets
contain in principle  different parameter degeneracies.  This can be seen in Figure~\ref{fig:sigma}:
the SG survey has a better handle on $\omega_c$, $N_\nu$ and $n_s$,
while G1 is more constraining for $h$.  Combination of these data sets, therefore,
can help lift these degeneracies (even if only partially).  Indeed, the combination of G1 and SG
yields a bound of $\sum m_\nu < 0.08 \ {\rm eV}$ (95 \% C.L.), which is
60 \% better than that from
G1 or SG alone.  A similar trend can also be seen in the allowed regions of $n_s$, further indicating that
we are indeed gaining better shape information from the combination.

This bound of $\sum m_\nu < 0.08 \ {\rm eV}$ from Planck+G1+SG is particularly noteworthy if 
we compare it with what can be achieved
in a low redshift survey with an effective volume equal to the combined volume of G1 and SG.
As an example, we might consider the survey ``G1$\times 1.45$'', which has $1.45$ times the
effective volume of G1.  From volume considerations
alone, the improvement in the parameter errors from G1$\times 1.45$ relative to G1
is at most a factor of $\sqrt{1.45} \sim 1.2$.  But this is a very optimistic figure because
the parameter constraints do not depend only on data from galaxy surveys; CMB data, too,
play a crucial role, and there is no corresponding increase in the CMB sky coverage.
Using an MCMC analysis, we find that Planck+G1$\times 1.45$ yields a
95 \% upper bound on $\sum m_\nu$ of 0.12 eV.  Compared with 0.13 eV from 
Planck+G1,
the actual improvement is negligible and cannot compete with the gain from
going to higher redshifts.

\begin{figure}
\hspace*{2.5cm}\includegraphics[width=130mm]{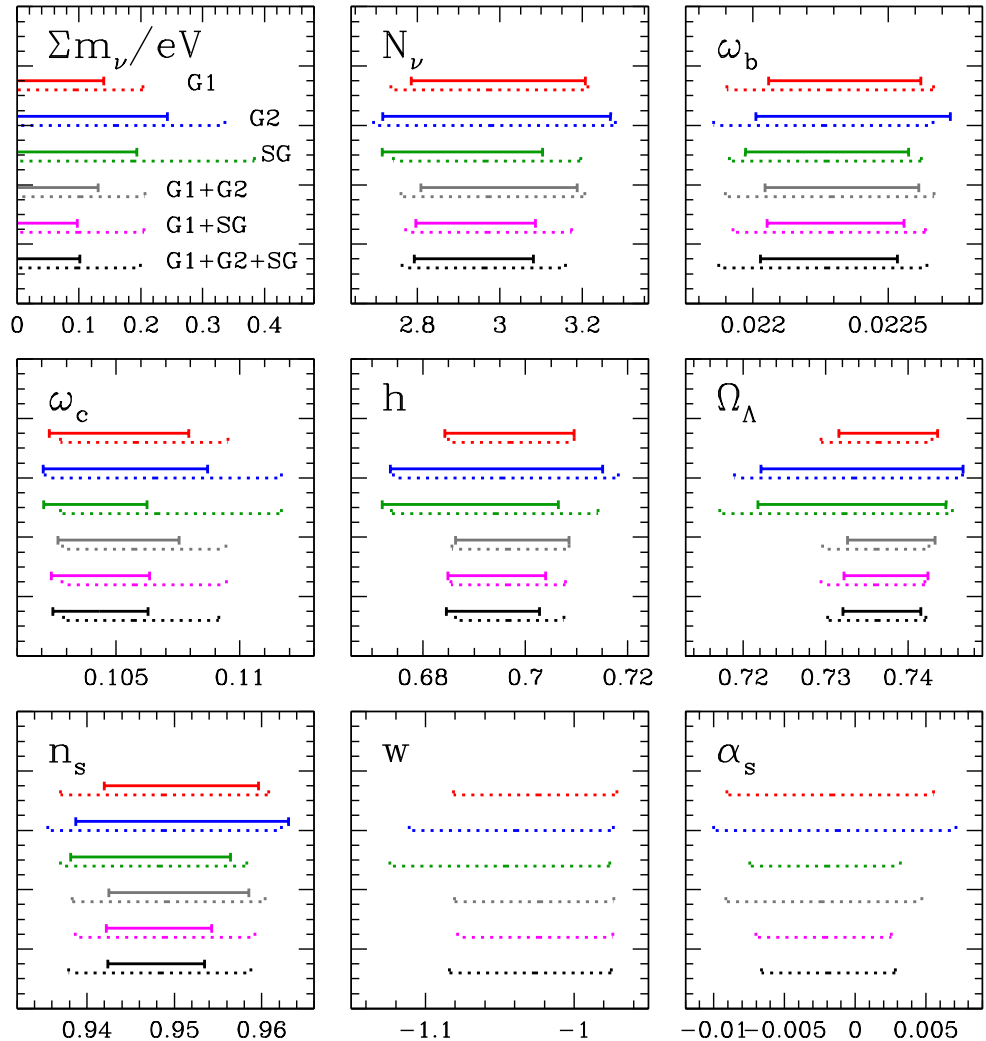} \caption{Same
as Figure~\ref{fig:sigma}, but with power spectrum smoothing as
described in section \ref{sec:windowfunctions}, i.e., no BAO
extraction.\label{fig:sigma_smooth}}
\end{figure}

For the 10-parameter model, however, the projected sensitivity to
$\sum m_\nu$ is much poorer and there is little gain even when both the
G2 and SG data sets are added to the analysis.
In this model the effect of the neutrino mass can be mimicked
by changes in a combination of other parameters, in this case mainly
the dark matter density $\omega_c$, the effective number of
neutrino species $N_\nu$, the dark energy equation of state $w$,
and the running of the spectral index $\alpha_s$. The larger $k$ range probed
by SG does lead to a better constraint on $\alpha_s$.  Unfortunately this
improvement does not translate to the neutrino mass bound.

We caution however that our treatment here, namely, the use of
spherically-averaged power spectra instead of two-dimensional spectra
with additional geometrical/redshift effects, may not be making full
use of the capacity of high-redshift galaxy surveys. Previous studies show
that the combination of geometrical effects and BAO can be a very
powerful tool to probe separately the Hubble expansion $H(z)$ and the
angular diameter distance $D_A(z)$ \cite{Seo:2003pu,Hu:2003ti}.
This may be useful for breaking the small degeneracy that exists between
$\sum m_\nu$ and $w$.

Figure~\ref{fig:sigma_smooth} shows
the projected 1D marginalised  95 \% confidence regions for the same
8- and 10-parameter models, but now using the smoothed spectra described in section
\ref{sec:windowfunctions}.  Compared with Figure~\ref{fig:sigma}, we see that
the inclusion of BAO information does lead to better constraints on
parameters in general.  However, it does not play a role in
improving in the neutrino mass bound  between Planck+G1+SG and
Planck+G1 in the 8-parameter model; the improvement seen in Figure~\ref{fig:sigma}
remains intact in Figure~\ref{fig:sigma_smooth} even without BAO extraction.
Clearly, the improved constraints come from tracing the evolution of the
LSS spectral shape.

Lastly, we note the 95 \% allowed ranges are in general not centred on the
fiducial model because of the non-Gaussian nature of the marginalised
posterior distribution.   The asymmetric error bars highlight
 the importance of using MCMC-based forecasts.

\begin{table}[t]
\caption{A summary of Figures~\ref{fig:sigma} and \ref{fig:sigma_smooth}:
projected 1D marginalised 90 \% (95 \%) upper bounds on $\sum m_\nu$, assuming
$\sum m_\nu^{\rm fid}=0$  (i.e., the ``sensitivity'').\label{table:sensitivity}}
\begin{indented}
\item[]\begin{tabular}{@{}lcccc}
\br
& \centre{2}{8-parameter model}  & \centre{2}{10-parameter model}   \\
\ns
& \crule{2} & \crule{2} \\
Survey & BAO & No BAO & BAO & No BAO \\
\br
Planck & \centre{2}{0.62 (0.73)}  & \centre{2}{0.78 (0.96)}  \\
\mr
+G1$\times$5 & 0.11 (0.13) & 0.12 (0.14)  & 0.17 (0.19) & 0.18 (0.20) \\
+G2 & 0.15 (0.18) & 0.20 (0.24) & 0.26 (0.30)  & 0.30 (0.34) \\
+SG & 0.12 (0.14) & 0.17 (0.19) & 0.30 (0.34)  & 0.34 (0.38) \\
+G1$\times$5+G2  & 0.10 (0.12) & 0.11 (0.13)  & 0.17 (0.19) & 0.18 (0.21) \\
+G1$\times$5+SG & 0.069 (0.080) & 0.086 (0.097) & 0.16 (0.18) & 0.18 (0.21) \\
+G1$\times$5+G2+SG & 0.069 (0.080) & 0.087 (0.10) & 0.16 (0.19) &  0.18 (0.20) \\
\br
\end{tabular}
\end{indented}
\end{table}

\subsection{Sensitivity}

It is interesting to compare the potential of galaxy redshift surveys
to constrain neutrino mass with the expected performance of
laboratory experiments such as
KATRIN.  The sensitivity of KATRIN, defined
as the 90 \% upper limit assuming a fiducial neutrino mass of zero, is
estimated to be $0.2 \ {\rm eV}$ \cite{Drexlin:2005zt}.  Derived under the same assumption of
$\sum m_\nu^{\rm fid} = 0$, Table~\ref{table:sensitivity} summarises
the projected 90 \% and 95 \%
sensitivities for various combinations of Planck and mock galaxy
surveys, with and without BAO extraction, for both the 8- and 10-parameter
models.

\begin{table}[t]
\caption{Projected 1D marginalised  95 \% lower and upper bounds
on $\sum m_\nu$ using Planck+G1+SG with BAO extraction  for
various fiducial $\sum m_\nu$ values.\label{table:threshold}}
\begin{indented}
\item[]\begin{tabular}{@{}lcccc}
\br
&  \centre{2}{8 parameters}  & \centre{2}{10 parameters}   \\
\ns
& \crule{2} & \crule{2} \\
$\sum m_\nu^{\rm fid}$ [eV] & Lower & Upper  & Lower & Upper \\
\br
0.0 &  0.0 & 0.080  & 0.0 & 0.19  \\
0.050 & 0.0 & 0.11 &  0.0 & 0.18 \\
0.093 & 0.027 & 0.14  & 0.0 & 0.21\\
0.16  & 0.11 & 0.20  & 0.0 & 0.26 \\
0.19  & 0.14 & 0.22  & 0.10 & 0.28 \\
0.22  & 0.18 & 0.26  & 0.14 & 0.31 \\
0.25  & 0.21 & 0.28  & 0.17 & 0.33 \\
0.28  & 0.23 & 0.31  & 0.21 & 0.36 \\
0.37  & 0.32 & 0.40  & 0.30 & 0.44 \\
\br
\end{tabular}
\end{indented}
\end{table}

\subsection{Detection threshold}

In addition to the sensitivity, KATRIN quotes a detection threshold,
defined as the fiducial neutrino mass at which an $N \sigma$ detection is
possible. With its present configuration, KATRIN will achieve a 3$\sigma$ detection
of an effective electron neutrino mass of 0.3 eV and $5 \sigma$ for 0.35 eV
\cite{Drexlin:2005zt}.

\begin{figure}
\hspace*{1.5cm}\includegraphics[width=100mm]{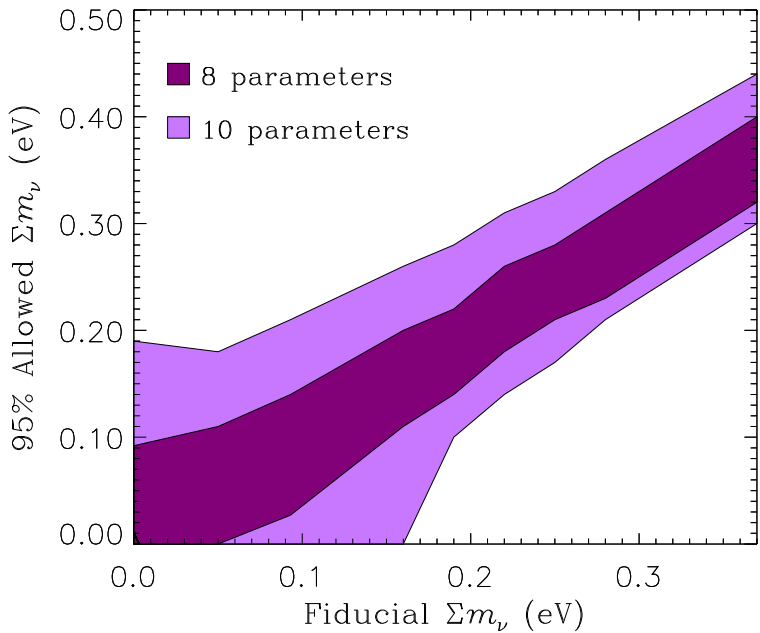} \caption{The
projected 1D marginalised 95 \% confidence range for $\sum m_\nu$
as a function of the fiducial $\sum m_\nu$ for Planck+G1+G2+SG with
BAO extraction. The light shaded band denotes the 10-parameter
model, while the dark band corresponds to the 8-parameter model.}
\label{fig:fig2}
\end{figure}

Calculating such a threshold for cosmological data is more involved because
it necessitates a scan in fiducial model space. Table~\ref{table:threshold}
and Figure~\ref{fig:fig2} show the results of such a scan; we derive the
1D marginalised 95 \% confidence regions using different values of $\sum
m_\nu^{\rm fid}$.  The projected 95 \% detection threshold is then defined
as the value of $\sum m_\nu^{\rm fid}$ at which the 95 \% lower bound differs
from zero.%
\footnote{One-tail limits are calculated when the marginalised posterior $P$ drops
below 15 \%  of its maximum at $\sum  m_\nu=0$.  This corresponds to
$-2 \ln P/P_{\rm max}  \sim  4$, i.e., roughly $2 \sigma$ in the case of a Gaussian
distribution.}
In the 8-parameter case the threshold thus defined
lies between 0.05 eV and 0.09 eV,
whereas in the the 10-parameter model it falls in the range 0.16 eV to 0.19 eV
because of the degeneracies between $\sum m_\nu$, $\alpha_s$ and $w$.
Comparing these numbers with their corresponding 95 \%
sensitivities derived in the last section (0.08 eV and 0.19 eV respectively),
we see that they are very similar, as one would intuitively expect.

Let us now consider what might be expected from a forecast based on the
Fisher matrix, which assumes Gaussianity in the posterior distribution with respect
to the model parameters.  There are two approaches---and two pitfalls.

The first approach is to derive $\sigma$ as function of $\sum m_\nu$, ignoring the fact
that the posterior distribution may be non-Gaussian for certain values of
$\sum m_\nu$.  Besides numerical instabilities
(see \cite{Perotto:2006rj} for a discussion), the $\sigma$ values derived in this
manner, and hence any confidence regions constructed therefrom, have no
meaning in the Bayesian context, simply because the posterior distribution is non-Gaussian.

The second approach is to avoid the non-Gaussian region by
evaluating the Fisher matrix at a larger fiducial value of $\sum
m_\nu$ at which the distribution is two-tailed and (hopefully)
approximately Gaussian.  The result is then extrapolated  to other
regions of parameter space. For example, in the case of the
10-parameter model, one might want to evaluate the Fisher matrix at,
say, $\sum m_\nu=0.25 \ {\rm eV}$. Granting that we can overcome the
issue of numerical instability, we should find a ``$2 \sigma$'' of
approximately 0.08 eV at $\sum m_\nu^{\rm fid}=0.25 \ {\rm eV}$ (see
Table~\ref{table:threshold}). However, the curvature of the
likelihood function is much larger for large values of the fiducial
neutrino mass, so that any such estimate of the detection threshold will likely be 
much too optimistic.
Indeed, extrapolating $2 \sigma=0.08$ eV
to lower values of $\sum m_\nu^{\rm fid}$, one would be led to conclude that the
detection threshold is 0.08 eV, which is clearly much lower than
($0.16 \to 0.19$) eV!

This highlights the importance of using MCMC techniques on
mock data instead of the Fisher matrix approach in order to get
reliable estimates of the potential of future experiments.

\section{Conclusions \label{sec:conclusions}}

We have calculated the sensitivity of future CMB probes and LSS
measurements from galaxy redshift surveys to
the neutrino mass. In particular, we have studied how the measurement
of the LSS power spectrum at different redshifts can help to
constrain $\sum m_\nu$ more efficiently than a single, more precise
measurement at low redshifts in some minimal cosmological models.

Along the same line we also comment on the difference between the
sensitivity, the formal upper limit given an underlying neutrino mass of
zero, and the detection threshold, the minimum $\sum m_\nu$
measurable at a given confidence level. We have quantified the
difference between the sensitivity and the detection threshold
in specific cases of CMB+LSS measurements.
We suggest that this method is the most
reliable for making error forecasts in cosmology. Furthermore, it has
the advantage that the results can be directly compared with projected
sensitivities and detection thresholds of laboratory experiments such as
KATRIN that probe directly the absolute neutrino mass scale.

Within the context of a minimal 8-parameter cosmological model,
we find that a minimum neutrino mass of order $0.05 \to 0.09$ eV can be detected at
95 \% confidence level using Planck and a combination of low and high redshift
galaxy surveys, while the 95 \% sensitivity is 0.08 eV.
The latest neutrino oscillation data
prefer  the mass splittings  $\Delta m_{12}^2
\sim 8 \times 10^{-5}$ eV$^2$ and $\Delta m_{23}^2 \sim 2.5 \times
10^{-3}$ eV$^2$
\cite{Smirnov:2007pw,Maltoni:2004ei}. If the
lightest eigenstate has zero mass, these mass splittings imply
$\sum m_\nu \simeq 0.06$ eV
for the normal hierarchy and 0.1 eV for the inverted hierarchy.
With the survey configurations chosen for our study, it will not be
possible to differentiate definitively between the two hierarchy schemes.

In more complex cosmological models with more free parameters, 
the detection threshold and sensitivity become even worse.  
In a 10-parameter model including also a running scalar spectral index and 
a nontrivial equation of state for the dark energy as free parameters,
Table~\ref{table:threshold} shows that
the sensitivity to $\sum m_\nu$
worsens by a more than a factor of two compared to the 8-parameter case.
The corresponding increase in
the detection threshold is threefold so that
a positive detection will only be possible
if neutrino masses are in the quasi-degenerate regime.

We note that the survey configurations employed in our analysis are not
necessarily overly optimistic. A very recent study of
neutrino mass measurements from Planck plus a future wide-field galaxy
survey out to $z \sim 2$ \cite{Abdalla:2007ut} assumes a much larger effective
survey volume than our G1 survey (roughly comparable to G1$\times 12$).
We have checked explicitly that the 95 \% sensitivity to $\sum m_\nu$
in this case is $0.06$ eV for the 8-parameter model, a figure that can
almost be matched by our 0.08 eV from a combination of low- and
high-redshift measurements with a much smaller total survey volume.

Finally we stress that although future high-redshift galaxy surveys
will be extremely useful for probing small neutrino masses, they are
by no means the only source of information about the LSS power spectrum. Weak
gravitational lensing of distant galaxies, for example, will provide
a complementary and possibly even more sensitive probe of neutrino properties
\cite{Cooray:1999rv,Abazajian:2002ck,Song:2003gg,Hannestad:2006as}.
In the more distant future, surveys of the 21-cm line from beyond the epoch
of reionisation will provide access to the matter power spectrum at even higher redshifts.

\section*{Acknowledgements}

We acknowledge computing resources from the Danish Center for
Scientific Computing (DCSC). SH acknowledges support from the
Alexander von Humboldt Foundation through a Friedrich Wilhelm Bessel
Award, and thanks the Max-Planck-Institut f\"ur Physik for
hospitality during the initial stages of this work. YYYW thanks Jan Hamann for
useful discussions and suggestions.


\section*{References}

\begin{thebibliography}{99}

\bibitem{Goobar:2006xz}
  A.~Goobar, S.~Hannestad, E.~Mortsell and H.~Tu,
  JCAP {\bf 0606} (2006) 019
  [arXiv:astro-ph/0602155].

\bibitem{Seljak:2006bg}
  U.~Seljak, A.~Slosar and P.~McDonald,
  JCAP {\bf 0610} (2006) 014
  [arXiv:astro-ph/0604335].

\bibitem{Feng:2006zj}
  B.~Feng, J.~Q.~Xia, J.~Yokoyama, X.~Zhang and G.~B.~Zhao,
  JCAP {\bf 0612} (2006) 011
  [arXiv:astro-ph/0605742].

\bibitem{Cirelli:2006kt}
  M.~Cirelli and A.~Strumia,
  JCAP {\bf 0612} (2006) 013
  [arXiv:astro-ph/0607086].

\bibitem{Hannestad:2006mi}
  S.~Hannestad and G.~G.~Raffelt,
  JCAP {\bf 0611} (2006) 016
  [arXiv:astro-ph/0607101].


\bibitem{Fogli:2006yq}
  G.~L.~Fogli {\it et al.},
 Phys.\ Rev.\ D {\bf 75} (2007) 053001 
  [arXiv:hep-ph/0608060].

\bibitem{Spergel:2006hy}
  D.~N.~Spergel {\it et al.},
Astrophys.\ J.\ Suppl.\ {\bf 170} (2007) 377
  [arXiv:astro-ph/0603449].


\bibitem{Tegmark:2006az}
  M.~Tegmark {\it et al.},
  Phys.\ Rev.\  D {\bf 74} (2006) 123507
  [arXiv:astro-ph/0608632].


\bibitem{Zunckel:2006mt}
  C.~Zunckel and P.~G.~Ferreira,
  arXiv:astro-ph/0610597.


\bibitem{Kristiansen:2006ky}
  J.~R.~Kristiansen, O.~Elgaroy and H.~Dahle,
  arXiv:astro-ph/0611761.

\bibitem{Lesgourgues:2006nd}
  J.~Lesgourgues and S.~Pastor,
  Phys.\ Rept.\  {\bf 429} (2006) 307
  [arXiv:astro-ph/0603494].


\bibitem{Cooray:1999rv}
  A.~R.~Cooray,
  Astron.\ Astrophys.\  {\bf 348}  (1999) 31
  [arXiv:astro-ph/9904246].

\bibitem{Abazajian:2002ck}
  K.~N.~Abazajian and S.~Dodelson,
  Phys.\ Rev.\ Lett.\  {\bf 91} (2003) 041301
  [arXiv:astro-ph/0212216].

\bibitem{Song:2003gg}
  Y.~S.~Song and L.~Knox,
  Phys.\ Rev.\ D {\bf 70} (2004) 063510
  [arXiv:astro-ph/0312175].

\bibitem{Hannestad:2006as}
  S.~Hannestad, H.~Tu and Y.~Y.~Y.~Wong,
  JCAP {\bf 0606} (2006) 025
  [arXiv:astro-ph/0603019].



%
\bibitem{Hu:1997mj}
  W.~Hu, D.~J.~Eisenstein and M.~Tegmark,
  Phys.\ Rev.\ Lett.\  {\bf 80} (1998) 5255
  [arXiv:astro-ph/9712057].

\bibitem{Hannestad:2002cn}
  S.~Hannestad,
  Phys.\ Rev.\  D {\bf 67} (2003) 085017
  [arXiv:astro-ph/0211106].

\bibitem{Lesgourgues:2004ps}
  J.~Lesgourgues, S.~Pastor and L.~Perotto,
  Phys.\ Rev.\  D {\bf 70} (2004) 045016
  [arXiv:hep-ph/0403296].

\bibitem{Takada:2005si}
  M.~Takada, E.~Komatsu and T.~Futamase,
  Phys.\ Rev.\ D {\bf 73} (2006) 083520
  [arXiv:astro-ph/0512374].


\bibitem{Glazebrook:2005ui}
  K.~Glazebrook, D.~Eisenstein, A.~Dey and B.~Nichol,
  arXiv:astro-ph/0507457.

\bibitem{bib:hetdex}
G.~J.~Hill, K.~Gebhardt, E.~Komatsu and P.~J.~MacQueen,
AIP Conf.\ Proc.\ {\bf 743} (2004) 224.

\bibitem{bib:cip}
{\tt http://www.cfa.harvard.edu/cip/}

\bibitem{Perotto:2006rj}
  L.~Perotto, J.~Lesgourgues, S.~Hannestad, H.~Tu and Y.~Y.~Y.~Wong,
  JCAP {\bf 0610} (2006) 013
  [arXiv:astro-ph/0606227].


\bibitem{Seo:2003pu}
  H.~J.~Seo and D.~J.~Eisenstein,
  Astrophys.\ J.\  {\bf 598} (2003) 720
  [arXiv:astro-ph/0307460].

\bibitem{Hu:2003ti}
  W.~Hu and Z.~Haiman,
  Phys.\ Rev.\  D {\bf 68} (2003) 063004
  [arXiv:astro-ph/0306053].

\bibitem{Hannestad:2005gj}
  S.~Hannestad,
  Phys.\ Rev.\ Lett.\  {\bf 95} (2005) 221301
  [arXiv:astro-ph/0505551].


\bibitem{Lewis:2002ah}
  A.~Lewis and S.~Bridle,
  Phys.\ Rev.\ D {\bf 66} (2002) 103511
  [arXiv:astro-ph/0205436].

\bibitem{cosmomc}{\tt http://cosmologist.info}

\bibitem{Tegmark:1997rp}
  M.~Tegmark,
  Phys.\ Rev.\ Lett.\  {\bf 79} (1997) 3806
  [arXiv:astro-ph/9706198].



\bibitem{Percival:2006gs}
  W.~J.~Percival {\it et al.},
Astrophys.\ J.\ {\bf 657} (2007) 51 
  [arXiv:astro-ph/0608635].

\bibitem{bib:planck}
{\tt
 http://astro.estec.esa.nl/SA-general/Projects/Planck/}

\bibitem{bib:bluebook}
{\tt http://www.rssd.eas.int/Planck}

\bibitem{Drexlin:2005zt}
  G.~Drexlin  [KATRIN Collaboration],
  Nucl.\ Phys.\ Proc.\ Suppl.\  {\bf 145} (2005) 263.



%
\bibitem{Smirnov:2007pw}
  A.~Y.~Smirnov,
  arXiv:hep-ph/0702061.

\bibitem{Maltoni:2004ei}
  M.~Maltoni, T.~Schwetz, M.~A.~Tortola and J.~W.~F.~Valle,
  New J.\ Phys.\  {\bf 6}  (2004) 122
  [arXiv:hep-ph/0405172].

\bibitem{Abdalla:2007ut}
  F.~B.~Abdalla and S.~Rawlings,
  arXiv:astro-ph/0702314.



\end{thebibliography}
\end{document}